\documentclass[11pt,a4paper]{article}

\usepackage{amssymb,epsfig}
\usepackage{amsmath}
\usepackage{amsthm}
\usepackage{amsfonts}
\usepackage{amssymb}
\usepackage{dsfont}
\usepackage{upref}
\usepackage{amsopn}
\usepackage{bm}

\usepackage{color}

\usepackage{epstopdf}

\usepackage{cite}

\usepackage{multirow}

\usepackage{nccmath}

\usepackage{slashed}
\usepackage[active]{srcltx}

\newcommand \widebar [1] {\overline{#1}}

\newcommand{\ms}{\mskip 1.5mu}

\newcommand \bigvev [1] {\big\langle{#1}\big\rangle}

\begin{document}

\begin{flushright}
DESY 18-175
\end{flushright}

\begin{center}
\textbf{\LARGE Conformal symmetry of QCD in $d$-dimensions}

\vspace*{1.2cm}

{\large V.~M.~Braun$\ms{}^1$,\ \ A.~N.~Manashov$\ms{}^{2,1,3}$,\  \ S.-O.~Moch$\ms{}^2$ and  M. Strohmaier$\ms{}^1$.}

\vspace*{0.4cm}

\textit{%
$^1$ Institut f\"ur Theoretische Physik, Universit\"at Regensburg,
D-93040 Regensburg, Germany \\
$^2$ II. Institut f\"ur Theoretische Physik, Universit\"at Hamburg,
   D-22761 Hamburg, Germany\\
$^3$ St.Petersburg Department of Steklov
Mathematical Institute, 191023 St.Petersburg, Russia} \vspace*{0.8cm}

\textbf{Abstract}\\[10pt]
\parbox[t]{0.9\textwidth}{
QCD in $d=4-2\epsilon$  space-time dimensions possesses a nontrivial critical point. Scale invariance  usually implies conformal
symmetry so that there are good reasons to expect that QCD at the critical point restricted to the gauge invariant subsector
provides one with an example of a conformal field theory. The aim of this letter is to present a technical proof of this statement
which is important both as a matter of principle and for applications.}
\end{center}

\vspace*{1cm}

\noindent{\large\bf 1.}~
Coupling constants in quantum field theory (QFT) models usually depend on the renormalization scale.
This dependence is described by beta-functions which enter
renormalization group equations (RGEs) for correlators of the fundamental fields
and/or local composite operators. If the beta-functions vanish, the theory enjoys scale invariance and the RGEs reduce to equations describing the behavior of the correlation functions under scale transformations.
In four-dimensional models, the only zero of the beta-functions accessible in perturbation theory corresponds to a trivial situation when all couplings vanish, i.e. the free theory.
In non-integer $d=4-2\epsilon$ dimensions, the situation is different. In this case it is common that the beta-functions vanish for some special values of the couplings $g = \mathcal{O}(\epsilon)$ (critical couplings).
If  $\epsilon$ is considered a small parameter, the critical couplings can be calculated in perturbation theory.

 QFT models at the critical point thus provide one with examples of  scale-invariant theories.
 As was first suggested by Polyakov~\cite{Polyakov:1970xd}, scale invariance
 usually implies conformal invariance. In particular it is believed that conformal symmetry follows unavoidably  from scale invariance if the theory is unitary, see refs.~\cite{Nakayama:2013is,Dymarsky:2013pqa,Dymarsky:2015jia} for the
 detailed argumentation. QFT models in non-integer dimensions are, however, not unitary~\cite{Hogervorst:2015akt} so that the proof does not apply.
 The question whether $d$-dimensional models at the critical point are conformal can, nevertheless, be answered, at least within
 perturbation theory, from the study of scale and conformal Ward identities.
 For non-gauge theories, conformal invariance of the correlators of
 fundamental fields can be proven along the lines of Refs.~\cite{Polchinski:1987dy,Vasilev1992,Derkachov:1993uw}.
 A detailed description of this technique and its extension
 to the case of local composite operators can be found in the book~\cite{vasil2004field}.

The situation with gauge theories and in particular QCD is more complicated. The gauge-fixing and ghost terms in the Lagrangian are not invariant under conformal transformations even in $d=4$ dimensions. As a consequence, there is no hope that
correlators of fundamental fields may transform in a proper way under scale and conformal transformations ---  good symmetry properties  can only be expected for the correlators of gauge-invariant operators. The subtlety is that
gauge-invariant operators mix under renormalization with gauge-variant operators of a special type (BRST variations) and Equation of Motion
operators (EOMs). These counterterms --- BRST and EOM operators --- are believed to be artifacts of the Faddeev-Popov approach to
quantization of gauge theories and all troubles caused by them are likely to be of technical character.
In this letter we clarify the structure of such "unwanted" contributions in conformal Ward identities, which is important
for practical applications. This analysis can be viewed as an extension of the work by Joglekar
and Lee~\cite{Joglekar:1975nu,Joglekar:1976eb,Joglekar:1976pe} on the structure of gauge-variant operators in the RGE equations.

It has been observed, see e.g.~\cite{Broadhurst:1993ru,Crewther:1997ux,Hatta:2008st,Caron-Huot:2015bja,Vladimirov:2016dll,Vladimirov:2017ksc},
that apparently unrelated perturbative QCD observables differ only by terms involving the beta-function,
and one possibility to understand this connection~\cite{Braun:2013tva,Vladimirov:2016dll,Vladimirov:2017ksc} is
to start from the theory in $d=4-2\epsilon$ dimensions at the critical point where they are related by a conformal transformation.
Similar ideas have been used to derive the RGEs for leading-twist QCD operators in general off-forward
kinematics \cite{Mueller:1997ak,Belitsky:1998gc,Braun:2016qlg,Braun:2017cih}.
Our intention is to put these methods on a more rigorous footing.

On a more technical level,
let $\mathcal{O}_q$, $q = 1,2,\ldots,n$ be a (finite) set of local composite operators with the same quantum numbers so that they
mix under renormalization.
In Ref.~\cite{vasil2004field} it was shown that in scalar theory the scale and conformal Ward identities for these operators
at the critical point imply that the symmetry transformations take the following form:
\begin{align}\label{DKgamma}
\delta_D \mathcal{O}_q(x) & =\left(D_{\Delta}(x)\delta_{qq'}+\gamma_{qq'} \right) \mathcal{O}_{q'}(x),
\notag\\
\delta_{K^\mu}\mathcal{O}_q(x) & = \left(K^\mu_{\Delta}(x)\delta_{qq'}+2\gamma_{qq'} x^\mu \right) \mathcal{O}_{q'}(x) +
\mathcal{O}^\mu_q(x),
\end{align}
where the sum over $q'$ is implied.  The generators of scale and conformal transformations $D_\Delta$ and $K^\mu_\Delta$
are defined as
\begin{align}
D_\Delta(x)=x\partial_x+\Delta, &&
K^\mu_\Delta(x)=
2x^\mu (x\partial)-x^2\partial^\mu + 2\Delta x^\mu -2x_\nu \Sigma^{\mu\nu},
\end{align}
where $\Delta$ is the canonical scaling dimension of the operators $\mathcal{O}_q$,
$\Sigma^{\mu\nu}$ is the spin generator and $\mathcal{O}^\mu_q$ are certain local operators with canonical dimension $\Delta-1$.
These expressions can be simplified by going over to a basis of  operators
that diagonalize the anomalous dimension matrix $\gamma$, $\mathcal{O}_q\longmapsto \mathcal{O}_{\Delta_\alpha}=c_{\alpha\,q}\mathcal{O}_q$.
Here $c_{\alpha q}$ is a left eigenvector of $\gamma_{qq'}$,
$\sum_q c_{\alpha q} \gamma_{qq'}=\gamma_\alpha c_{\alpha q'}$, and $\Delta_\alpha=\Delta+\gamma_\alpha$ is the scaling dimension of
the operator $\mathcal{O}_{\Delta_\alpha}$.
In this basis the transformations in Eqs.~\eqref{DKgamma} simplify to
\begin{align}\label{DKalpha}
\delta_D \mathcal{O}_{\Delta_\alpha} (x) = D_{\Delta_\alpha}(x)\mathcal{O}_{\Delta_\alpha} (x), &&
\delta_{K^\mu} \mathcal{O}_{\Delta_\alpha} (x) = K^\mu_{\Delta_\alpha}(x)\mathcal{O}_{\Delta_\alpha} (x) +\mathcal{O}^\mu_\alpha(x),
\end{align}
where $\mathcal{O}^\mu_\alpha=c_{\alpha q} \mathcal{O}^\mu_q$. Scale invariance implies that the operator
$\mathcal{O}^\mu_\alpha$ has definite scaling dimension equal to $\Delta_\alpha -1$. The set of operators with the same
anomalous dimensions (meaning that the difference of  scaling dimensions of any two operators is an integer number), forms an
infinite-dimensional representation (Verma module) of the conformal algebra. The expressions in Eqs.~\eqref{DKalpha} define the action of scale and conformal generators on
this  representation. Since the scaling dimension of the operator $\mathcal{O}_\alpha^\mu$ is less than that of
$\mathcal{O}_{\Delta_\alpha}$ by one,  applying the conformal transformations subsequently to
$\mathcal{O}_{\Delta_\alpha}$, $\mathcal{O}_\alpha^\mu$ etc. one inevitably must come to an operator
for which the addendum  $\mathcal{O}_\alpha^\mu$ on the r.h.s.
vanishes, i.e. an operator that transforms homogeneously under conformal transformations. Such an operator is called conformal and it is the lowest weight vector of the corresponding representation.

The analysis of scale and conformal Ward identities given in Ref.~\cite{vasil2004field} can be extended to gauge theories.
We will show that  Eqs.~\eqref{DKalpha} keep their form.
The main result is that the inhomogeneous part, $\mathcal{O}^\mu_\alpha(x)$, in the expression for the conformal variation of a
gauge-invariant operator is a gauge invariant operator again, up to terms that vanish in all correlation functions of gauge-invariant operators and can therefore always be dropped.

At first sight the appearance of a gauge non-invariant operator on the r.h.s. of Eqs.~\eqref{DKalpha} can be ruled out by observing
that its anomalous dimension would depend on the gauge-fixing parameter. This is not always the case, however.
To give an example, the gauge-invariant operator
$\mathcal{O}=F \widetilde F$ in four dimensions can be written as a divergence of the topological current $K^\mu$,
$F \widetilde F=\partial_\mu K^\mu$.  Evidently, $F \widetilde F$ and $K^\mu$ have the same anomalous dimensions and the
current $K^\mu$ can be a natural candidate for the role of the non-homogeneous  term  in Eqs.~\eqref{DKalpha},
$\delta_{K^\mu} \mathcal{O}(0)\sim K^\mu(0)$. At the same time $K^\mu$ is not a gauge-invariant operator.

\vspace*{0.2cm}

\noindent{\large\bf 2.}~ We start with collecting the necessary definitions. The QCD action in $d=4-2\epsilon$ Euclidean space
reads
\begin{align}\label{SQCD}
S=\int d^dx \Big\{\bar q \slashed{D} q+\frac14 F_{\mu \nu}^aF^{a,\mu\nu}
- \bar c^a\partial_\mu(D^\mu c)^a +\frac1{2\xi}(\partial_\mu A^{a,\mu})^2\Big\}\,,
\end{align}
where
$D_\mu={\partial_\mu}-ig_B A_\mu^a T^a$  with $T^a$ being the $SU(N_c)$ generators in the fundamental (adjoint)
representation for quarks (ghosts). The field strength tensor is defined as usual,
$F_{\mu\nu}^a=\partial_\mu A_\nu^a -\partial_\nu A_\mu^a +g_B f^{abc}A_\mu^b A_\nu^c$,
where $g_B$ is the bare coupling, $g_B=g M^\epsilon$, and  $M$ is the scale parameter. The theory is assumed to be multiplicatively
renormalized and the renormalized action takes the form $S_R(\Phi,e)=S(\Phi_0,e_0)$, where $\Phi=\{A,q,\bar q,c,\bar c\}$,
$e=\{g,\xi\}$ and
$\Phi_0=Z_\Phi \Phi$, $e_0=Z_e e$. The renormalization factors in the minimal subtraction (MS) scheme have a series expansion in $1/\epsilon$,
\begin{align}
 Z = 1 + \sum\limits_{j=1}^\infty \epsilon^{-j} \sum\limits_{k=j}^\infty z_{jk}\,  a^k\,,
\qquad a ={\alpha_s}/{(4\pi)} = {g^2}/{(4\pi)^2}\,,
\end{align}
where $z_{jk}$ are polynomials in $\xi$. Formally the theory has two charges: $a$ and $\xi$. The corresponding beta-functions
are defined as
\begin{align}\label{}
\beta_a(a) = M\frac{d g}{dM} = 2a\big(-\epsilon - \gamma_g\big)\,,
&&
\beta_\xi(\xi,a) = M\frac{d\xi}{dM} = -2\xi\gamma_A(a,\xi)\,,
\end{align}
with
\begin{align}
\gamma_g=M\partial_M \ln Z_g = \beta_0\, a
+ \beta_1\,a^2
+ \mathcal{O}(a^3)\,,
\end{align}
where the first two coefficients are $\beta_0 = {11}/{3} N_c - 2/3 N_f$, $\beta_1=2/3\left[ 17 N_c^2-5N_c N_f -3 C_F N_f\right]$
for a $SU(N_c)$ gauge group with $N_f$ quark flavors.
The anomalous dimensions of the fields $ \Phi=\{q,\bar q, A,c,\bar c \}$ are defined as
\begin{align}
\gamma_\Phi = M\partial_M \ln Z_\Phi=\big(\beta_g \partial_g +\beta_\xi\partial_\xi\big)\ln Z_\Phi \,.
\end{align}

The QCD Lagrangian~\eqref{SQCD} is invariant under BRST transformations~\cite{Becchi:1975nq,Tyutin:1975qk},
$\delta\mathcal{L}=0$, where
\begin{align}
\label{BRSTtrafos}
\delta q & =ig t^a q c^a\delta\lambda\,,
&
\delta A_\mu^a  &= (D_\mu c)^a\delta\lambda\,,
&
\delta c^a & =\frac12 g f^{abc} c^b c^c\delta\lambda\,, & \delta \bar c^a & =-\frac1\xi (\partial A^a)\delta \lambda\,.
\end{align}
The BRST transformation rules for the renormalized fields  are obtained by replacement $\Phi\mapsto \Phi_0$, $e \to e_0$, $\delta
\lambda\to \delta\lambda_0$ in the above equations and writing the bare fields and couplings in terms of the renormalized ones:
$\Phi_0=Z_\Phi \Phi$, $e_0 = Z_e e$. The renormalized BRST transformation parameter $\delta\lambda$ is defined as $\delta\lambda_0=Z_c Z_A \delta\lambda$ so that the last equation in Eqs.~\eqref{BRSTtrafos}
has the same form for bare and renormalized quantities. The BRST operator $s$ defined by $\delta \Phi = s\,\Phi\,\delta\lambda$ is
nilpotent  modulo  EOM terms. Namely,  $s^2\Phi=0$ for all fields except for the anti-ghost in which case one finds
\begin{align}
s^2 \bar c=-\frac1\xi s (\partial A)= -\frac1\xi Z_c^2 (\partial^\mu D_\mu) c=
\frac1\xi \frac{\delta S_R}{\delta \bar c}.
\end{align}
Thus the second BRST variation of an arbitrary local functional $\mathcal{F}(\Phi)$ is an EOM operator
\begin{align}\label{s2F}
s^2\mathcal{F}(\Phi)=\frac1\xi \int d^dx \frac{\delta S_R}{\delta \bar c^a(x)}\frac{\delta \mathcal{F}(\Phi)}{\delta \bar c^a(x)}\,.
\end{align}

BRST symmetry is the key ingredient in the analysis of the RGEs for gauge invariant operators~\cite{Joglekar:1975nu,Joglekar:1976eb,Joglekar:1976pe}.
The result, see Ref.~\cite{collins_1984} for a review,
is that gauge invariant operators, $\mathcal{O}$, mix under renormalization with BRST operators, i.e. operators that can be written as a BRST variation of another operator, $\mathcal{B}=s \mathcal{B}'$, and EOM
operators, $\mathcal{E}=F(\Phi)\delta S_R/\delta\Phi$. The mixing matrix has a triangular structure
\begin{align}
\begin{bmatrix}
\mathcal{O} \\
\mathcal{B} \\ \mathcal{E}
\end{bmatrix}
=   \begin{pmatrix}
Z_{\mathcal{OO}} &Z_\mathcal{OB}& Z_{\mathcal{OE}}\\
 0&Z_\mathcal{BB}& Z_{\mathcal{BE}}\\
0 & 0& Z_{\mathcal{EE}}
\end{pmatrix}
\begin{pmatrix}
\mathcal{O} \\
\mathcal{B} \\ \mathcal{E}
\end{pmatrix},
\end{align}
so that renormalized gauge-invariant operators take the following generic form
 \begin{align}
[\mathcal{O}] =Z_{\mathcal{OO}}\mathcal{O} + Z_{\mathcal{OB}} \mathcal{B}+Z_\mathcal{OE} \mathcal{E}
\equiv \widehat{\mathcal{O}} + Z_{\mathcal{OB}} \mathcal{B}+Z_\mathcal{OE} \mathcal{E}
\,,
 \end{align}
 where we introduced a notation $\widehat{\mathcal{O}}=Z_{\mathcal{OO}}\mathcal{O}$ for the gauge-invariant part of
 the renormalized (gauge-invariant) operator. Note that the renormalization factor $Z_\mathcal{OO}$ does not
 depend on the gauge parameter $\xi$.

It should in principle be possible to constrain the operator structure of
potential BRST and EOM counterterms for a given $\mathcal{O}$.
However, no such relation is known.

The significance of this result is that the contributions of BRST and EOM operators to physical
observables have to vanish so that such terms can be dropped, at least in principle.
In practice this requires some caution. Calculations are usually done in momentum space.
Within perturbation theory the radiative corrections to the matrix elements of composite operators
develop ultra-violet divergences as well as of infrared ones, which are regularized in $d$ dimensions.
In addition, the vanishing of physical matrix elements with BRST or EOM operators
requires the on-shell limit with respect to their external momentum $q$ to be taken and,
generally, the limits $q^2\to 0$ and $d\to 4$ do not commute.
Therefore, theorems on the renormalization of gauge invariant operators~\cite{Joglekar:1975nu,Joglekar:1976eb,Joglekar:1976pe}
directly apply to matrix elements with the operators inserted at nonzero momentum.
In practice, this requires the computation of three-point functions with
off-shell legs, which poses certain difficulties at higher loops.
Calculations of matrix elements based on two-point functions are technically easier, but
are typically realized with operators inserted at zero momentum.
In this case, physical matrix elements of gauge variant operators do not vanish, the
mixing matrix of operators is not triangular and matrix elements with insertions of BRST
or EOM operators need to be accounted for as well,
see refs.~\cite{Hamberg:1991qt,Collins:1994ee,Harris:1994tp,Matiounine:1998ky}.

Considering operators with fixed position essentially corresponds to nonzero momentum flow. In this case it is indeed
easy to see that a correlation function
of renormalized gauge-invariant operators localized at different space-time points $x_k$ can be written as
\begin{align}\label{OhatnohatO}
\bigvev{\prod_k [\mathcal{O}_{k}(x_k)]}=
    \bigvev{ \prod_k \widehat{\mathcal{O}}_{k}(x_k)} + \sum_{k,m} \delta(x_k-x_m)C_{km}(\vec{x})=
        \bigvev{ \prod_k \widehat{\mathcal{O}}_{k}(x_k)}.
\end{align}
Our goal in this paper is to show that at the critical point, $\beta_a(a_*)=0$, the correlators~\eqref{OhatnohatO}
behave in a proper way under scale and conformal transformations.
The last expression in the above identity is a natural starting point for this undertaking.

\vspace*{0.2cm}

\noindent{\large\bf 3.}~
Next, we introduce the relevant Ward identities.
The correlation function in Eq.~\eqref{OhatnohatO} can be written in the path-integral representation as follows
\begin{align}\label{FIO}
 \bigvev{ \prod_k \widehat{\mathcal{O}}_{k}(x_k)}=\mathcal{N} \int D\Phi \,\prod_k \widehat{\mathcal{O}}_{k}(x_k)\,
  \exp\big\{-S_R(\Phi)\big\},
\end{align}
where $\mathcal{N}$ is the normalization factor. Making the change of variables $\Phi\mapsto \Phi'=\Phi+\delta_\omega \Phi$ in the
integral~\eqref{FIO}, where $\delta_\omega\Phi$ correspond to the dilatation and special conformal transformation, $\omega=D,  K^\mu$,
see Appendix~\ref{app:SK}, and taking into account
that the integration measure stays invariant, one obtains
\begin{align}\label{WI}
\sum_j \bigvev{\delta_\omega \widehat{\mathcal{O}}_j(x_j) \prod_{k\neq j} \widehat{\mathcal{O}}_k(x_k)}=
 \bigvev{\delta_\omega S_R\prod_{k} \widehat{\mathcal{O}}_k(x_k)}.
\end{align}
Note the choice of the canonical dimensions for the fields in Eq.~\eqref{canonical-dimensions}.
For this choice the commutator of dilatation/conformal $\delta_\omega$ and gauge transformations $\delta_\alpha$ is a gauge transformation again,
\begin{align}
\label{omegaalpha}
[\delta_\alpha,\delta_{\omega}]&=\delta_{\alpha_\omega}\,,
\end{align}
where $\alpha_D=(x\partial)\alpha$ and $\alpha_{K^\mu}=\Big (2x^\mu(x\partial)-x^2\partial^\mu\Big)\alpha$.

Assuming that the operators $\widehat{\mathcal{O}}_j$ have canonical dimensions $\Delta_j$ one finds for their variations that appear
on the l.h.s. of Eq.~\eqref{WI},
$\delta_\omega \mathcal{O}(x)= \int d^d y\, \delta_\omega\Phi(y) \big({\delta\mathcal{O}(x)}/{\delta\Phi(y)}\big)$,
the following expressions:
\begin{align}\label{SKOG}
\delta_D \widehat{\mathcal{O}}_j(x) = D_{\Delta_j} \widehat{\mathcal{O}}_j(x)\,, &&
\delta_{K^\mu}\widehat{\mathcal{O}}_j(x) = K^{\mu}_{\Delta_j} \widehat{\mathcal{O}}_j(x) + \sum_k p_{jk}\widehat{\mathcal{O}}_k^\mu(x),
\end{align}
where $\widehat{\mathcal{O}}_k^\mu$ are certain gauge invariant operators with canonical dimension $\Delta_j-1$.
Such inhomogeneous terms typically arise from the commutators of $\delta_w$ with derivatives in the operator
$\mathcal{O}_j$, if they are present.
Note that the coefficients $p_{jk}(\epsilon)$ can be and, as a rule, are singular in the $\epsilon \to 0$ limit.
It is easy to check that the property \eqref{omegaalpha} ensures that there are no
gauge-dependent addenda to these expressions.

The variation of the QCD action $\delta_\omega S_R$ on the r.h.s. of the Ward identity \eqref{WI}, see Appendix,
can be written as
\begin{align}
\label{SKS}
\delta_D S_R  =\int d^dx \,2\epsilon\,\mathcal{L}'_R(x)\,,
&&
\delta_{K^\mu} S_R = \int d^dx\,\Big(4\epsilon\, x^\mu\mathcal{L}'_R(x)
-2(d-2)\partial^\rho [\mathcal{B}_{\rho}](x)\Big)\,,
\end{align}
where $\mathcal{L}'_R(x)=\mathcal{L}_R(x)-\tfrac12 Z_q^2\, \partial^\rho \,\big(\bar q(x)\gamma_\rho q(x)\big) $ and
$\mathcal{B}_{\rho}$ is a BRST operator, see Eq.~\eqref{Brho}. This term does not contribute to the correlation function,
$\bigvev{ \mathcal{B}_{\rho}(x)\prod_{k} \widehat{\mathcal{O}}_k(x_k)}=0 $, so that the r.h.s. of Eq.~\eqref{WI}
takes the standard form
\begin{align}\label{hatCWI}
2\epsilon \int d^dx \, \chi_\omega(x) \bigvev{\mathcal{L}_R'(x)\prod_{k} \widehat{\mathcal{O}}_k(x_k)},
\end{align}
where $\chi_D=1$ and $\chi_{K^\mu}=2x^\mu$ for dilatation and conformal transformations, respectively.

To proceed further we re-expand
$2\epsilon \mathcal{L}'_R(x)$ in terms of renormalized (finite) operators.
The corresponding expression takes the form~\cite{Spiridonov:1984br,Belitsky:1998gc,Braun:2016qlg}
\begin{align}\label{LQCDR}
{2\epsilon}\mathcal{L}'_R&=-\frac{\beta(a)}{a}\left[\mathcal{L}^{YM}+\mathcal{L}^{gf}\right]-(\gamma_q-\epsilon) \Omega_{q\bar q}
-(\gamma_A+\gamma_g)\Omega_A-(\gamma_c-2\epsilon)\Omega_{\bar c} -\gamma_c\Omega_c+2\gamma_A [\mathcal{L}^{gf}]
\notag\\
&\quad
+z_b(g,\xi) \partial_\mu [\mathcal{B}^\mu]+z_c(g,\xi)\partial_\mu [\Omega^\mu].
\end{align}
Here $\Omega_\Phi =\Phi(x){\delta S_R}/{\delta \Phi(x)}$, $\Omega_{q\bar q}=\Omega_q+\Omega_{\bar q}$ and
$\Omega_\mu= \bar c D_\mu c-\partial_\mu \bar c \,c$ is a conserved current,
$\partial_\mu [\Omega^\mu]=\Omega_c-\Omega_{\bar c}$.
  The gauge fixing term
 $\mathcal{L}^{gf}=\frac1{2\xi} (\partial A)^2$ can be rewritten as a combination of BRST and EOM operators,
\begin{align}
[\mathcal{L}^{gf}]=-[\mathcal{B}]-\Omega_{\bar c}, &&    \mathcal{B}=s\big(\bar c^a\,(\partial A^a)\big)\,.
\end{align}
It can be shown that the  coefficients $z_b(g,\xi)$ and $z_c(g,\xi)$ can be calculated explicitly in Landau gauge,
$\xi=0$,
\begin{align}
z_b(g,\xi)=\gamma_A+\gamma_g +O(\xi), && z_c(g,\xi)=-(\gamma_A+\gamma_g)/2+O(\xi)\,.
\end{align}
Using Eq.~\eqref{LQCDR} in Eq.~\eqref{hatCWI} it is easy to see that only the contributions coming from small integration regions
around the points $x_k$ survive at the critical point. Indeed, let $B_k$ be an arbitrary small ball centered at $x_k$ and split the
integration region in two parts: the union of the (non-overlapping) small balls $B=\bigcup_k B_k$ and their complement $\widebar R
=R^d\backslash B$. Integrating over the complement one can drop all EOM terms appearing in Eq.~\eqref{LQCDR} and also the
contributions of the BRST operators. Thus this contribution reduces to
\begin{align}\label{hatbarB}
-\frac{\beta(a)}a \int_{\widebar R} d^dx \, \chi_\omega(x) \bigvev{[\mathcal{L}^{YM}(x)]\prod_{k} \widehat{\mathcal{O}}_k(x_k)}.
\end{align}
The remaining correlation function contains renormalized (finite) local operators at separated space points and is
finite. The integral is also finite.
This contribution vanishes, therefore, at the critical point since it comes with the factor
$\beta(a_*)=0$. Thus only the integral over the union of small balls around the operator insertions remains,
\begin{align}\label{hatB}
2\epsilon \sum_n \int_{B_n} d^dx \, \chi_\omega(x) \bigvev{\mathcal{L}_R'(x)\prod_{k} \widehat{\mathcal{O}}_k(x_k)}.
\end{align}
Our next aim is to bring this expression to a form suitable for further analysis.

\vspace*{0.2cm}

\noindent{\large\bf 4.}~
Since the balls $B_n$ do not overlap, it is sufficient to consider one term in the sum.   The operator product
$2\epsilon\mathcal{L}'_R(x)\widehat{\mathcal O}_n(x_n)$ for $x\to x_n$ is not necessarily finite and
the argument which we used to claim that the integral over the complement $\widebar
R$ can be dropped does not work.
To simplify the notation we suppress the subscript $n$ and use  $x' \equiv x_n$.
The first step is to show that the product of the renormalized Lagrangian and a gauge-invariant renormalized operator $\widehat{\mathcal{O}}(x')$ can be written
in the following form
\begin{align}\label{eL-exp}
2\epsilon\mathcal{L}'_R(x)\widehat{\mathcal{O}}(x') %
& =
-\frac{\beta(a)}a [\mathcal{L}^{YM}(x) \mathcal{O}(x')]
+\mathrm{LT}(x,x') + s(R(x,x')) +\mathcal{E}(x,x').
\end{align}
The first term on the r.h.s. of this expression is the fully renormalized product of two operators.
$\mathrm{LT}$ stands for local terms that have a finite expansion of the form
\begin{align}\label{Lxxprime}
\mathrm{LT}(x,x')=\delta(x-x')\, \mathcal{F}(x')+\partial_x^\mu\delta(x-x')\, \mathcal{F}^\mu (x')+\ldots
\end{align}
The next term is a BRST operator.
Finally, the last term is an EOM operator which has the following property: its correlation function with a product of
fundamental fields
$\mathcal{X}(Y)=\prod_p \Phi_p(y_p)$, $Y = \{y_1,\ldots,y_p\}$ contains only delta functions of the type $\delta(x-y_p)$ or $\delta(x'-y_p)$
but not $\delta(x-x')$. In other words if $x,x' \neq y_p$ for any $p$ then
\begin{align}
\label{Exx'}
\bigvev{\mathcal{E}(x,x')\, \mathcal{X}(Y)} = 0.
\end{align}

In order to prove Eq.~\eqref{eL-exp} we start with the representation \eqref{LQCDR} for the QCD Lagrangian. This expression
contains several terms: EOM operators, BRST variations and the renormalized Yang-Mills part of the Lagrangian $[\mathcal{L}^{YM}]$
which comes with the factor $\beta(a)$. In what follows we examine these contributions one-by-one.
\begin{itemize}
\item It is straightforward to  show that the EOM terms give rise to
\begin{align}\label{ExO}
\mathcal{E}(x)\widehat{\mathcal{O}}(x')=\mathrm{LT}(x,x')+\mathcal{E}(x,x').
\end{align}
To this end consider the correlation function of $\mathcal{E}(x)\widehat{\mathcal{O}}(x')$ with a set of
fundamental fields $\mathcal{X}(Y)$ which we can write as
\begin{align}
    \bigvev{\Phi(x)\frac{\delta S_R}{\delta\Phi(x)} \widehat {\mathcal{O}}(x') \mathcal{X}} =
 \bigvev{\Phi(x)\frac{\delta \widehat {\mathcal{O}}(x') }{\delta\Phi(x)}\mathcal{X}} +
\bigvev{\Phi(x)\biggl(\frac{\delta S_R}{\delta\Phi(x)} \widehat {\mathcal{O}}(x') -\frac{\delta \widehat {\mathcal{O}}(x') }{\delta\Phi(x)}\biggr)\mathcal{X}}.
\end{align}
The first term on the r.h.s. is a local operator while the second one is a EOM term, $\mathcal{E}(x,x')$, that is easy to see
integrating by parts in the path integral.
\item The product  $\mathcal{B}(x)\widehat{\mathcal{O}}(x') $ can be written as $s(\mathcal{B}'\widehat{\mathcal{O}}(x'))$
and, therefore, contributes to the $R(x,x')$ term only.
\item The last term to consider is ${\beta(a)}/a \, [\mathcal{L}^{YM}]\, \widehat{\mathcal{O}}(x')$. Here we replace
$\widehat{\mathcal{O}}(x')$ by the complete renormalized operator $[{\mathcal{O}}(x')]$ and subtract the corresponding
BRST and EOM counterterms. The latter ones contribute to $\mathrm{LT}(x,x')$ and $\mathcal{E}(x,x')$, cf.
Eq.~\eqref{ExO}.  The product of two renormalized operators $[\mathcal{L}^{YM}(x)]$ and
$[\widehat{\mathcal{O}}(x')]$ can be written as a sum of the renormalized operator product and local pair counterterms,
\begin{align}
[\mathcal{L}^{YM}(x)][\widehat{\mathcal{O}}(x')]=[\mathcal{L}^{YM}(x)\widehat{\mathcal{O}}(x')] + \mathrm{LT}(x,x')\,.
\end{align}
We are left with the product of $[\mathcal{L}^{YM}](x)$ and the BRST counterterm to $[{\mathcal{O}}(x')]$, call it $\mathcal{B}_O(x')$. Separating the gauge-invariant part
\begin{align}
 [\mathcal{L}^{YM}](x) = \widehat{\mathcal{L}}^{YM}_L +\mathcal{B}_L(x) + \mathcal{E}_L
\end{align}
we observe that the EOM term gives rise to the structure~\eqref{ExO} whereas the product $\widehat{\mathcal{L}}^{YM}_L\mathcal{B}_O(x')$ contributes to the $R(x,x')$ term. Finally, the product of two BRST operators $\mathcal{B}_L(x)=s(\mathcal{B}'_L(x))$ and $\mathcal{B}_O(x')=s(\mathcal{B}'_O(x'))$ can be
rewritten as
\begin{align}
\mathcal{B}_L(x)\mathcal{B}_O(x')=s\big(\mathcal{B}'_L(x)\mathcal{B}_O(x')\big)-\mathcal{B}'_L(x)s\big(\mathcal{B}_O(x')\big).
\end{align}
The first term on the r.h.s. contributes to $R(x,x')$ and the the second term is the sum of local (LT) and
$\mathcal{E}(x,x')$ (EOM) contributions. To see this, write
 $s\big(\mathcal{B}_O(x')\big)=s^2\big(\mathcal{B}'_O(x')\big)$ and
use Eq.~\eqref{s2F} to obtain
\begin{align}
\mathcal{B}'_L(x)s\Big(\mathcal{B}'_O(x')\Big)=
\mathcal{B}'_L(x)\frac1\xi \int d^dz \frac{\delta S_R}{\delta \bar c^a(z)}\frac{\delta \mathcal{B}'_O(x')}{\delta \bar c^a(z)}
=\frac1\xi\int d^dz \frac{\delta\mathcal{B}'_L(x)}{\delta \bar c^a(z)}\frac{\delta \mathcal{B}'_O(x')}{\delta \bar c^a(z)}+ \mathcal{E}(x,x').
\end{align}
Obviously, the first term on the r.h.s. of this identity is a local (LT) contribution.
Collecting all of the above expressions we obtain Eq.~\eqref{eL-exp}.
\end{itemize}

Once Eq.~\eqref{eL-exp} is established, we can use it in the correlation function~\eqref{hatB}.
The EOM term $\mathcal{E}(x,x')$ drops out thanks to Eq.~\eqref{Exx'} and the BRST operator $sR(x,x')$ obviously does not
contribute as well. The first term,
$-\tfrac{\beta(a)}a [\mathcal{L}^{YM}(x) \mathcal{O}(x')]$, vanishes at the critical point. Thus the sole contribution
to the correlation function~\eqref{hatB} at the critical point is due to the local terms, $\mathrm{LT}(x,x')$.
As seen from the above analysis the local terms originate from different sources and
separate contributions are clearly gauge  non-invariant. Nevertheless, it is possible to show that the complete expression
for $\mathrm{LT}(x,x')$ can be written as a sum of the contributions of gauge-invariant, BRST and EOM operators.

The proof follows closely the analysis of the RGEs for gauge-invariant operators in Ref.~\cite{collins_1984}.
To this end we consider the BRST variation of Eq.~\eqref{eL-exp}.
Since the l.h.s. vanishes, one obtains
\begin{align}\label{seL}
s\big(\mathrm{LT}(x,x')\big)=\frac{\beta(a)}{a} s\big( [\mathcal{L}^{YM}(x) \mathcal{O}(x')]\big)
-s^2\big(R(x,x')\big) - s\big(\mathcal{E}(x,x')\big).
\end{align}
Using
\begin{align}
\bigvev{s\big(\mathcal{E}(x,x')\big)\,\mathcal{X}}=-\bigvev{\mathcal{E}(x,x')\, s(\mathcal{X})}\,,
    &&
\bigvev{s^2\big(R(x,x')\big)\,\mathcal{X}}=\bigvev{R(x,x')\,s^2(\mathcal{X})}\,,
\end{align}
where, as above,  $\mathcal{X}=\prod_p \Phi_p(y_p)$ and $x,x'\neq y_p$, it is easy to see that
the last two terms in Eq.~\eqref{seL} are EOM operators, $\mathcal{E}(x,x')$.
Next, we want to show that
$s\big( [\mathcal{L}^{YM}(x) \mathcal{O}(x')]\big)$ is an EOM operator as well.
The starting observation is that BRST variations of the fundamental fields are finite operators~\cite{collins_1984} and therefore
the BRST variation of a renormalized operator is a finite operator as well, up
to EOM operators. Using the same arguments that lead to Eq.~\eqref{eL-exp} one can show that
for a product of any two gauge-invariant operators one gets
\begin{align}
[\mathcal{O}_1(x)\mathcal{O}_2(x')]=\widehat{\mathcal{O}}_1(x)\widehat{\mathcal{O}}_2(x') + \mathrm{LT}(x,x') +s(R(x,x'))
+\mathcal{E}(x,x')\,,
\end{align}
where all terms on the r.h.s. except for the first one are singular in $1/\epsilon$ (do not contain finite contributions).
Taking a BRST variation of the both sides we conclude that up to EOM terms
$
s([\mathcal{O}_1(x)\mathcal{O}_2(x')])= s(\mathrm{LT}(x,x')).
$
The operator on the l.h.s. of this relation is a finite operator, while the one on the r.h.s. is singular. Therefore they both
are equal to zero, up to EOM terms.

Going back to Eq.~\eqref{seL} we conclude that $s(\mathrm{LT}(x,x'))= 0$ modulo EOM operators. As shown by Joglekar
and Lee~\cite{Joglekar:1975nu}, see also~\cite{Henneaux:1993jn} for a review, vanishing of the BRST variation implies
that $\mathrm{LT}(x,x')$ and therefore the operators $\mathcal{F},\mathcal{F}^\mu$
in Eq.~\eqref{Lxxprime} can be written as a sum of gauge invariant, BRST and EOM operators.
The last ones can safely be neglected since they do not contribute to the correlation function in question.

\vspace*{0.2cm}

\noindent{\large\bf 5.}~
The subsequent derivation of the scale and conformal properties of correlation functions of gauge-invariant operators
follows the lines of Ref.~\cite{vasil2004field}. Starting from the dilatation Ward identity in Eq.~\eqref{WI}
and taking into account Eqs.~\eqref{SKOG},~\eqref{hatB},~\eqref{eL-exp} one obtains
\begin{align}\label{sWI}
\sum_j \bigvev{\left( D_{\Delta_j}(x_j) \widehat{\mathcal{O}}_j(x_j) -\mathcal{F}_j(x_j)
	\right)\prod_{k\neq j} \widehat{\mathcal{O}}_k(x_k)} = 0\,.
\end{align}
Taking into account that the operators in questions satisfy the RGEs
\begin{align}
M\partial_M [\mathcal{O}_k] +\sum_{k'}\gamma_{kk'} [\mathcal{O}_{k'}]=0\,,
\end{align}
and have definite canonical dimension
\begin{align}
\Big(M\partial_M -\sum_j D_{\Delta_j}(x_j)\Big)\bigvev{\prod_{k} \widehat{\mathcal{O}}_k(x_k)}=0\,,
\end{align}
this identity implies that~\footnote{Our notations are a bit sloppy here. The sum over $j'$ goes over all operators which mix
	with $\mathcal{O}_j(x_j)$. We do not assume that the operators at different points belong to the same class.}

\begin{align}\label{sWI-1}
\sum_j \bigvev{\Big( \sum_{j'}\gamma_{jj'} \widehat{\mathcal{O}}_{j'}(x_j) +\mathcal{F}_j(x_j)
	\Big)\prod_{k\neq j} \widehat{\mathcal{O}}_k(x_k)} = 0\,.
\end{align}
Since this equation must hold for arbitrary operator insertions $\prod_{k\neq j} \widehat{\mathcal{O}}_k(x_k)$ one concludes that
\begin{align}
\label{Fdilatation}
\mathcal{F}_j(x_j)= - \sum_{j'}\gamma_{jj'} \widehat{\mathcal{O}}_{j'}(x_j).
\end{align}
The same relation can alternatively be achieved  by the analysis
of the dilatation Ward identity for the correlation function of local operators with fundamental fields in Landau gauge.
In this gauge $\beta_\xi=0$ holds identically so that the both beta-functions vanish at the critical point and scale invariance
holds for any Green's function.

Using Eq.~\eqref{Fdilatation} we can rewrite the conformal Ward identity as follows:
\begin{align}
\sum_j \bigvev{\left( K^\mu_{\Delta_j}(x_j) \widehat{\mathcal{O}}_j(x_j) - 2x^\mu \mathcal{F}_j(x_j) + \widetilde{\mathcal{F}}_j^\mu(x_j)
	\right)\prod_{k\neq j} \widehat{\mathcal{O}}_k(x_k)} = 0\,,
\end{align}
where $ \widetilde{\mathcal{F}}_j^\mu(x)= 2{\mathcal{F}}_j^\mu(x)-\sum_k p_{jk}\widehat{\mathcal{O}}_k^\mu(x) $, see
Eq.~\eqref{SKOG}. Note that all divergent terms in $\widetilde{\mathcal{F}}_j^\mu(x)$ have to cancel.

Finally, using Eq.~\eqref{SKOG}, we obtain
\begin{align}
\delta_D \widehat{\mathcal{O}}_j(x) & = \Big(\delta_{jj'}D_{\Delta_j}(x_j)+\gamma_{jj'}\Big) \widehat{\mathcal{O}}_{j'}(x)\,,
\notag\\
\delta_{K^\mu} \widehat{\mathcal{O}}_j(x) & = \Big(\delta_{jj'}K_{\Delta_j}(x_j)+2 x^\mu \gamma_{jj'}\Big) \widehat{\mathcal{O}}_{j'}(x)
+\widehat{\mathcal{O}}^\mu_j(x)\,,
\end{align}
where $\widehat{\mathcal{O}}^\mu_j(x)=\widetilde{\mathcal{F}}_j^\mu(x)$ is a gauge-invariant operator and the operator
equality holds up to terms that vanish for all correlation functions with any number of gauge-invariant operators.
 Provided that the anomalous dimension matrix can be diagonalized~\footnote{This is not always possible in theories with fermions
 where the number of mixing operators can be infinite, see~Ref.~\cite{Ji:2018emi}.} one can go over to the basis of operators with definite scaling dimensions and rewrite these equations in the form~\eqref{DKalpha}.

\vspace*{0.2cm}

\noindent{\large\bf 6.}~
To summarize, we have shown by the BRST analysis of the corresponding Ward identities  that correlation functions of gauge-invariant
operators in QCD in $d=4-2\epsilon$ dimensions at the critical point transform properly under conformal transformations, as expected
in a conformal invariant theory. This result gives further support to the methods based on using conformal invariance in
higher-order perturbative QCD calculations~\cite{Broadhurst:1993ru,Crewther:1997ux,Mueller:1997ak,Belitsky:1998gc,Hatta:2008st,Caron-Huot:2015bja,Vladimirov:2016dll,Vladimirov:2017ksc,Braun:2016qlg,Braun:2017cih}
and can be also interesting in a broader context.

\vspace*{0.2cm}

\noindent{\bf Acknowledgments}\\
We thank Yu.~Pismak for a useful discussion.
The work by AM  was supported by the DFG grant MO~1801/1-3 and the RSF project 14-11-00598.

\appendix
\renewcommand{\theequation}{A.\arabic{equation}}
\renewcommand{\thesection}{{\Alph{section}}}
\renewcommand{\thetable}{\Alph{table}}
\setcounter{section}{0} \setcounter{table}{0}
\setcounter{equation}{0}

\section*{Appendix: Scale and conformal transformations}\label{app:SK}

The dilatation (scale) $D$ and conformal $K$ transformations  for the fundamental fields take the form
\begin{align}\label{SandK}
\delta_D \Phi(x)
    &=D_{\Delta_\Phi}(x) \Phi(x) =
    \big(x\partial_x+\Delta_\Phi\big) \Phi(x),
\notag\\
\delta_{K^\mu} \Phi(x)
    &=K^\mu_{\Delta_\Phi}(x) \Phi(x) =
    \big(2x^\mu(x\partial)-x^2\partial^\mu +2\Delta_\Phi x^\mu -2x_\nu \Sigma^{\mu\nu}\big) \Phi(x)\,,
\end{align}
in particular
\begin{align}\label{}
K_\mu q(x)
    &=\big(2x_\mu(x\partial)-x^2\partial_\mu +2\Delta_q\, x_\mu\big)\, q(x)+\frac12[\gamma_\mu,\slashed{x}]q(x),
\notag\\
K_\mu \bar q(x)
    &=\big(2x_\mu(x\partial)-x^2\partial_\mu +2\Delta_q\, x_\mu\big)\, \bar q(x)-\bar q(x)\frac12[\gamma_\mu,\slashed{x}],
\notag\\
K_\mu c(x)
    &=\big(2x_\mu(x\partial)-x^2\partial_\mu +2\Delta_c\, x_\mu\big)\, c(x),
\notag\\
K_\mu \bar c(x)
    &=\big(2x_\mu(x\partial)-x^2\partial_\mu +2\Delta_{\bar c}\, x_\mu\big)\, \bar c(x),
\notag\\
K_\mu A_\rho(x)
    &=\big(2x_\mu(x\partial)-x^2\partial_\mu +2\Delta_A\, x_\mu\big)\, A_\rho(x) + 2g_{\mu\rho} (x A)-2x_\rho A_\mu(x),
\end{align}
where $\Delta_\Phi=\dim \Phi$ are the field canonical dimensions. It is convenient to choose them in $d=4-2\epsilon$ dimensions
to be the same as in four-dimensional theory,
\begin{align}\label{canonical-dimensions}
\Delta_A=1, && \Delta_{q}=\Delta_{\bar q}=3/2, && \Delta_c=0, &&\Delta_{\bar c}=2.
\end{align}
For this choice  the field strength tensor $F_{\sigma\rho}$ transforms in a covariant way 
\begin{align}\label{F-transform}
K_\mu F_{\sigma\rho} &=\Big(2x_\mu(x\partial)-x^2\partial_\mu +4 x_\mu\Big)F_{\sigma\rho}
+2\Big(g_{\mu\rho} x^\nu F_{\sigma\nu}+g_{\mu\sigma} x^\nu F_{\nu\rho} -x_\rho F_{\sigma\mu}-x_\sigma F_{\mu\rho}\Big), &
\end{align}
and the covariant derivative of the ghost field $D_\nu c$ transform as a vector field,
\begin{align}\label{}
K_\mu D_\rho c(x) &=\big(2x_\mu(x\partial)-x^2\partial_\mu +2x_\mu\big) D_\rho c(x)
+2\big(g_{\mu\rho} (x D)-x_\rho D_\mu\Big)c(x)\,. &
\end{align}

A conformal variation  of different pieces of the QCD action takes the form
\begin{subequations}
\begin{align}
\label{Kqq}
\delta_K\int d^dx \bar q \slashed{D} q &=
    4\epsilon \int d^dx \Big(x^\mu \bar q \slashed{D} q~+~\frac12 \bar q \gamma_\mu q\Big),
\\
\label{F2}
\delta_K\int d^dx \frac14 F^2 &=
    4\epsilon \int d^dx \,x^\mu\frac14 F^2\,,
\\
\label{KGFt}
\delta_K\int d^dx\frac1{2\xi}(\partial A)^2 &=
    -\frac1\xi\int d^dx\Big(-2\epsilon\, x^\mu (\partial A)^2+2(d-2) A^\mu (\partial A)\Big)\,,
\\
\label{Sghost}
\delta_K\int d^dx\Big(-\bar c\partial_\mu D^\mu c\Big) & =
    4\epsilon \int d^dx x^\mu\Big(-\bar c\partial_\mu D^\mu c\Big) +2(d-2)\int d^dx \, \bar c D^\mu c.
\end{align}
\end{subequations}
Note that the ghost and the gauge fixing terms break the conformal symmetry  explicitly even in $d=4$ dimensions. Summing up all
contributions yields
\begin{align}
\label{DS}
\delta_D S & =\int d^dx \,2\epsilon\mathcal{L}(x)\,,   \\
\label{KS}
\delta_{K^\mu} S &= \int d^dx\,\biggl(4\epsilon\, x^\mu\Big( \mathcal{L}(x) -\frac12 \partial^\rho
\mathcal{J}_\rho(x)\Big)-2(d-2)\partial^\rho \mathcal{B}_{\rho}(x)\biggr)\,.
\end{align}
Here $\mathcal{J}_{\rho}(x)=\bar q(x) \gamma_\rho q(x)$ is the flavor-singlet vector current and
\begin{flalign}
\label{Brho}
\mathcal{B}_\mu=\bar c D_\mu c -\frac1\xi A_\mu (\partial A)
\end{flalign}
is a BRST operator, $\mathcal{B}_\mu =s(\bar c A_\mu)$.

\bibliography{new_conf-3}

\end{document}